\documentclass[letterpaper, 10 pt, conference]{ieeeconf}  

\IEEEoverridecommandlockouts                              

\usepackage{graphics} 
\usepackage{epsfig} 
\usepackage{mathptmx} 
\usepackage{times} 
\usepackage{amsmath} 
\usepackage{amssymb}  
\usepackage{mathtools}
\usepackage{tcolorbox}
\usepackage{pgfkeys}
\usepackage{xcolor}
\usepackage{pgfplots}
\usepackage{layouts}
\usepackage{tikz}
\usepackage{algorithm}
\usepackage[noend]{algpseudocode}
\usepackage{graphicx}
\usepackage[mathcal]{eucal}
\usepackage{microtype}

\newlength\figurewidth
\definecolor{new}{rgb}{0,0,0}

\title{\LARGE \bf
Sample-based Moving Horizon Estimation 
}

\author{Isabelle Krauss, Victor G. Lopez and Matthias A. Müller
	\thanks{This work received funding from the European Research Council (ERC) under the European Union’s Horizon 2020 research and innovation programme (grant agreement No 948679).}
	\thanks{I. Krauss, V. G. Lopez and M. A. Müller are with the Leibniz University Hannover, Institute of Automatic Control, 30167 Hannover, Germany
		{\tt\small \{krauss,lopez,mueller\}@irt.uni-hannover.de}}%
}

\newcommand\copyrighttext{%
	\footnotesize \copyright 2026 IEEE. Personal use of this material is permitted. Permission from IEEE must be obtained for all other uses, in any current or future media, including reprinting/republishing this material for advertising or promotional purposes, creating new collective works, for resale or redistribution to servers or lists, or reuse of any copyrighted component of this work in other works.}
\newcommand\copyrightnotice{%
	\begin{tikzpicture}[remember picture,overlay]
		\node[anchor=south,yshift=7pt] at (current page.south) {\fbox{\parbox{\dimexpr\textwidth-\fboxsep-\fboxrule\relax}{\copyrighttext}}};
	\end{tikzpicture}%
}

\begin{document}
		\newtheorem{thm}{Theorem}
\newtheorem{cor}{Corollary}
\newtheorem{lem}{Lemma}
\newtheorem{prop}{Proposition}

\newtheorem{rem}{Remark}

\newtheorem{defi}{Definition}
\newtheorem{ass}{Assumption}

\setlength\figurewidth{0.9\columnwidth}

\maketitle
\thispagestyle{empty}
\pagestyle{empty}
\copyrightnotice

\begin{abstract}
	In this paper, we propose a sample-based moving horizon estimation (MHE) scheme for general nonlinear systems to estimate the current system state using irregularly and/or infrequently available measurements. The cost function of the MHE optimization problem is suitably designed to accommodate these irregular output sequences. We also establish that, under a suitable sample-based detectability condition known as sample-based incremental input/output-to-state stability \mbox{(i-IOSS)}, the proposed sample-based MHE achieves robust global exponential stability (RGES).
	Additionally, for the case of linear systems, we draw connections between sample-based observability and sample-based i-IOSS. 
	This demonstrates that previously established conditions for linear systems to be sample-based observable can be utilized to verify or design sampling strategies that satisfy the conditions to guarantee 
	RGES of the sample-based MHE. Finally, the effectiveness of the proposed sample-based MHE is illustrated through a simulation example.
\end{abstract}
\section{INTRODUCTION}
Moving Horizon Estimation (MHE) is an optimization-based method for estimating the internal state of a dynamical system. It formulates the estimation task as a constrained optimization problem over a finite horizon using a sequence of past output measurements and inputs.
MHE has proven to be a powerful solution to the state estimation problem, particularly  due to its ability to handle  general nonlinear systems,  potentially with state constraints, and subject to model inaccuracies and measurement noise \cite{Raw22}. 
Theoretical analysis of MHE has established strong stability results. In particular, robust stability can be guaranteed under a mild detectability assumption -- incremental input/output-to-state stability (i-IOSS) -- as shown in, e.g., \cite{All21,Ji16,Knu23,Sch23}. 
\par While standard MHE formulations rely on the assumption of constant, regular sampling, there are many applications in which only infrequent and/or irregular output measurements are available. This may result from practical constraints that make measuring the system outputs continuously or at every time instant infeasible. A representative example arises in the biomedical field, where measurements such as blood samples are collected only sparsely. These measurements are then for instance used to estimate internal physiological states -- such as hormone concentrations -- to e.g. diagnose disorders in the hypothalamic–pituitary–thyroid axis \cite{Die16} and to derive appropriate medication strategies (see, e.g., \cite{Bru21,Wol22}). Under such circumstances suitable state estimators are needed that can handle this limited output information to still recover the internal state. 
\par To design suitable state estimators under irregular or infrequent output sampling, it is essential to establish sample-based observability or detectability conditions that explicitly account for the limited availability of measurements. For linear systems, such conditions have been studied for continuous-time systems in \cite{Wan11,Zen16}, and for discrete-time systems in \cite{Kra22}. The nonlinear discrete-time case was addressed in \cite{Kra25}, where a sample-based version of incremental input/output-to-state stability (i‑IOSS) was introduced.
\color{new} Although \cite{Kra25} investigated detectability conditions for nonlinear systems with irregular measurement sequences, these results have not yet been exploited for state estimation.
\color{black}
\par Building upon these results, this paper proposes a sample-based formulation of MHE 
for state estimation in systems where output measurements are irregular and/or sparse.
In particular, our contributions are as follows. We propose a sample-based MHE scheme for general nonlinear discrete-time systems (Section~\ref{sec:MHEscheme}). We then establish robust global exponential stability (RGES) of the estimation error under a sample-based detectability assumption (Section~\ref{sec:stabana}). Furthermore, we demonstrate how sample-based observability conditions for linear systems relate to the sample-based i-IOSS framework proposed in \cite{Kra25} (Section~\ref{sec:LinSys}). Thereby, we show that existing conditions for sample-based observability in linear systems can be leveraged to verify or design sampling strategies that satisfy the sample-based i-IOSS condition, thus  ensuring RGES of the sample-based MHE. Finally, the effectiveness of the proposed scheme is demonstrated through a simulation example (Section~\ref{sec:numex}).
\section{PRELIMINARIES AND SETUP}
The set of all nonnegative real numbers is denoted by $\mathbb{R}_{\geq0}$, the set of integers greater than or equal to $a$  for some  $a\in \mathbb{R}$  by $\mathbb{I}_{\geq a}$, and the set of integers in the interval $[a,b]$  for some  \mbox{$a,b \in \mathbb{R}$}  by $\mathbb{I}_{[a,b]}$.  The bold symbol $\mathbf{u}$ refers to a sequence of the vector-valued variable \mbox{${u\in\mathbb{R}^m}, \ \mathbf{u}= \{u_0,u_1,\ldots\}$}  and the notation $(\mathbb{R}^m)^\infty$ denotes the set of all
sequences $\mathbf{u}$ with infinite length. $P\succ 0$ ($P\succeq 0$) denotes a positive (semi-)definite matrix.
We denote the euclidean norm of vector $x \in \mathbb{R}^n$   by $||x||$ and $||x||_P^2=x^\top Px$ with  $P=P^\top\succ 0$. 
Furthermore,  $\sigma_{\text{min}}(P)$ and  $\sigma_{\text{max}}(P)$ refer to the minimal and maximal singular value of $P$, respectively. Analogously, $\lambda_{\text{min}}(P)$ and  $\lambda_{\text{max}}(P)$ refer to the minimal and maximal eigenvalue of $P$, respectively, and  $\lambda_{\text{max}}(P,Q)$ denotes the  maximum generalized eigenvalue of positive definite matrices $P,Q$. 
\par We consider the discrete-time nonlinear system 
\begin{align}
	\begin{aligned}
		x_{t+1}&=f(x_t,u_t,w_t) \\
		y_t&=h(x_t,u_t,w_t)\\
	\end{aligned}
	\label{eq:sys}
\end{align}
with state $x \in \mathbb{X} \subseteq \mathbb{R}^n$, control input $u\in \mathbb{U} \subseteq \mathbb{R}^m$, disturbance $w \in \mathbb{W} \subseteq \mathbb{R}^q$ with $0 \in \mathbb{W}$,  noisy output measurement $y \in \mathbb{Y} \subseteq \mathbb{R}^p$, time $t \in \mathbb{I}_{\geq 0}$, and nonlinear continuous functions $f: \mathbb{X} \times \mathbb{U} \times \mathbb{W} \rightarrow \mathbb{X}, \ h: \mathbb{X} \times \mathbb{U} \times \mathbb{W}\rightarrow \mathbb{Y}$ representing the system dynamics and the output model, respectively. Notice that we use $w$ to denote both process and measurement noise. 
\par	A widely used notion of detectability for nonlinear \mbox{systems --} particularly  in the context of MHE -- is incremental input-output-to-state stability (i-IOSS), compare, e.g., \cite{Son95,All21,Ji16,Knu23,Sch23}.
\begin{defi}[i-IOSS]
	The system (\ref{eq:sys}) is 
	i-IOSS if there exist functions $\beta_x,\beta_w,\beta_y \in \mathcal{KL}$  such that for any two initial
	conditions $x_{0}$, $\tilde{x}_{0}$, any pair of input (disturbance) trajectories $\mathbf{w},\mathbf{\tilde{w}}\in\mathbb{W}^\infty$ and all $u\in\mathbb{U}^\infty$ 
	the following holds for all $t \geq 0$
	\begin{align}
		\begin{aligned}
			||x_t-\tilde{x}_t||
			&\leq \max\Big\{\beta_x(|| x_0-\tilde{x}_0
			||,t	), \\&\max_{j\in \mathbb{I}_{[0,t-1]}}\beta_w(||w_j-\tilde{w_j}||,t-j-1),\\
			& \max_{j\in \mathbb{I}_{[0,t-1]}}\beta_y(||y_j-\tilde{y}_j||,t-j-1)\Big\}.
		\end{aligned}
		\label{eq:td-iIOSS}
	\end{align}
	If additionally $\beta_x(r,t)=C_1 r  \lambda_1^t$, $\beta_w(r,t)=C_2 r \lambda_2^t$,  and  $\beta_y(r,t)=C_3 r \lambda_3^t$  with $\lambda_1,\lambda_3,\lambda_3 \in [0,1)$ and $C_1,C_2,C_3>0$ then the system  is  exponentially i-IOSS.
	\label{def:iioss}
\end{defi}
\par If (\ref{eq:td-iIOSS}) holds for some $\beta_x,\beta_w \in \mathcal{KL}$ and $\beta_y \equiv 0$, the system is said to be incrementally input-to-state stable (\mbox{i-ISS}) \cite{Ang02}.
\par In \cite{Kra25} a sample-based i-IOSS formulation was introduced in which irregular or infrequent measurement sequences are explicitly accounted for in the output-dependent term of the \mbox{i-IOSS} bound. 
The sample-based detectability condition relies on the following definition of  a set  $K$.
\begin{defi}[Sampling set $K$ \cite {Kra25}]
	\label{def:K}
	Consider an infinitely long sequence $\mathcal{D}=\{d_1,d_2,\ldots\}$ with $d_i \in \mathbb{I}_{\geq 0}, \ i\in \mathbb{I}_{> 0}$ and  $\max_i d_i \eqcolon d_{\text{max}}<\infty$. 
	\color{new}	The set $K_i=\{t_1^i,t_2^i,\ldots\}$ refers to an infinite set of time instances with 	$t_1^i=d_i, \ 	t_j^i=t_{j-1}^i+d_{i+j-1},  \forall j\in\mathbb{I}_{\geq2}$. \color{black}
	The set $K$ then refers to a set of sets containing all $K_i$, $ i\in\mathbb{I}_{> 0}$.
\end{defi}
\color{new}
Defining the sampling set $K$ in this way is useful for analyzing sample-based detectability uniformly in time. For a more detailed discussion, the reader is referred to  \cite{Kra25}.
\color{black}
\par In this paper, we propose a sample-based MHE for which robust exponential stability can be established. For this purpose, we use an exponential version of the sample-based \mbox{i-IOSS} notion introduced in \cite{Kra25} as the detectability assumption\footnote{\color{new} Note that, unlike the setting considered here, \cite{Kra25} does not include a control input $u$. Nevertheless, the results in \cite{Kra25} can be readily extended to incorporate $u$.}. 
\color{black}
\begin{ass}[sample-based exponential  i-IOSS]
Consider a set $K$ as in Definition~\ref{def:K}.
System (\ref{eq:sys}) is sample-based exponentially i-IOSS with respect to $K$, i.e., there exist  $P_1,P_2 \succ 0$, $Q, R \succeq 0$ and $\eta \in [0, 1)$ such that for any pair of disturbance trajectories  $\mathbf{w},\mathbf{\tilde{w}} \in\mathbb{W}^\infty$, any pair of initial conditions  $x_{0}$, $\tilde{x}_{0}$ and all $u\in\mathbb{U}^\infty$ it holds that for all $t\geq 0$ and any $K_i\in K$
\begin{align}
\begin{aligned}
	||x_t-\tilde{x}_t||_{P_1}^2&\leq ||x_0-\tilde{x}_0||_{P_2}^2 \eta^t +\sum_{j=0}^{t-1}\eta^{t-j-1}||w_j-\tilde{w_j}||_Q^2
	\\&+\sum_{j\in \mathbb{I}_{[0,t-1]} \cap K_i}\eta^{t-j-1}||y_j-\tilde{y}_j||_R^2.
\end{aligned}
\label{eq:ass1}
\end{align}
\label{ass:eIOSS}
\end{ass}
\par Assumption~\ref{ass:eIOSS} is such that only those measurements $y_j$ are considered in the second sum in (\ref{eq:ass1}) for which $j\in K_i$.  A sufficient condition for  Assumption~\ref{ass:eIOSS} to hold has been presented in \cite{Kra25} and will also be recalled in Section~\ref{sec:LinSys} (compare Theorem~\ref{thm1}). While this condition can be difficult to verify \emph{a priori} in the general nonlinear case, we also discuss in Section~\ref{sec:LinSys} that for the special case of linear systems, existing results on sample-based observability can be used to verify or design sampling schemes that satisfy Assumption~\ref{ass:eIOSS}.
\begin{rem}
\label{rem:eIOSS}
Note that in Assumption~\ref{ass:eIOSS}, a sum-based formulation of sample-based exponential i-IOSS was used, 
whereas a max-based version is typically used in the general (non-exponential) case, compare (\ref{eq:td-iIOSS}). In fact, in case of (sample-based) \emph{exponential}  i-IOSS, the two formulations are equivalent, compare \cite[Remarks 3 and 4]{Knu20}. We use the sum-based version in (\ref{eq:ass1}) as it facilitates some of our arguments in the sample-based MHE stability proof of Theorem~\ref{thm:stab}. 
\end{rem} 
\par 	In Section~\ref{sec:stabana}, we establish robust exponential stability of the scheme introduced in Section~\ref{sec:MHEscheme}, according to the following stability definition.
\begin{defi}[RGES]
\label{def:rges}
A state estimator for system (\ref{eq:sys})  providing the estimate $\hat{x}_t$ is 
RGES if there exist
$C_x,C_w>0$ and $\lambda_x,\lambda_w \in [0,1)$  such that for any initial conditions $x_0,\hat{x}_0 \in \mathbb{X}$ and any disturbance sequence~$\mathbf{w}\in\mathbb{W}^\infty$ the following holds for all $t\geq 0$
\begin{align*}
\begin{aligned}
	||x_t-\hat{x}_t||\leq C_x||x_0-\hat{x}_0||\lambda_x^t+\sum_{j=0}^{t-1}
	C_w||w_j||\lambda_w^{t-j-1}.
\end{aligned}
\label{eq:rges}
\end{align*} 
\end{defi}
\section{SAMPLE-BASED MHE}
\label{sec:MHEscheme}
MHE is an optimization-based state estimation technique that computes state estimates by solving an optimization problem over a sliding window of past inputs and outputs at each time instant $t$.
In this work, we propose a sample-based MHE scheme, that  only takes an irregular and/or infrequent measurement sequence into account. The sequence $K_s$ contains the time instances where a measurement is available to the estimator.
The horizon length is defined as $M_t\coloneqq\min\{t,M+\delta_t\}$ with $\delta_t:=t-1-\max\{0,j \in K_s|j<t\}$ referring to the time that has passed since the last measurement was available.
Importantly, in the proposed scheme, the optimization problem does not need to be explicitly solved at every time instant $t$ but only when new measurement information becomes available. At these times, by definition, $\delta_t=0$, so that for $t\geq M$ the nonlinear program (NLP) is solved with a fixed horizon length $M$. When no new information is available, i.e., $\delta_t>0$, then the optimal estimate is given by an open-loop prediction (cf. Proposition~\ref{prop:NLPOL} below).
Although the optimization problem is only solved when new measurements arrive, the following NLP is defined for all times $t$. Later, in the proof of Theorem~\ref{thm:stab}, we exploit the fact that the solution of this optimization problem exists at all times. Thus, using the time-varying horizon length $M_t$, consider the following problem
\begin{subequations}
\label{eq:NLP}
\begin{align}
\min_{\hat{x}_{t-M_t|t}, \hat{w}_{\cdot|t}} 
&	J(\hat{x}_{t-M_t|t}, \hat{w}_{\cdot|t},   \hat{y}_{\cdot|t},t)\\
\text{s.t.} \  \hat{x}_{j+1|t}&=f(\hat{x}_{j|t},u_j,\hat{w}_{j|t}), \ j\in \mathbb{I}_{[t-M_t,t-1]},\label{eq:NLPC1}\\
\hat{y}_{j |t}&=h(\hat{x}_{j|t},u_j,\hat{w}_{j|t}), \ j\in \mathbb{I}_{[t-M_t,t-1]}
,\label{eq:NLPC2}\\
\hat{w}_{j|t} &\in \mathbb{W}, \hat{y}_{j|t} \in \mathbb{Y}, \ j\in \mathbb{I}_{[t-M_t,t-1]},\label{eq:NLPC3}\\
\hat{x}_{j|t} &\in \mathbb{X}, \ j\in \mathbb{I}_{[t-M_t,t]}. \label{eq:NLPC4}
\end{align}
\end{subequations}
The notation $\hat{x}_{j|t}$ denotes the estimated state at time $j$ computed at the current time $t$. The notations  $\hat{w}_{j|t}$ and $\hat{y}_{j|t}$ for the estimated disturbances and outputs follow analogously.
The optimal state, disturbance, and output sequences that minimize the cost function $J$ are denoted by  $\hat{x}^*_{\cdot|t}$, $\hat{w}^*_{\cdot|t}$ and  $\hat{y}^*_{\cdot|t}$, respectively.
The optimal estimate at the current time $t$ is indicated by $\hat{x}_t \coloneqq \hat{x}^*_{t|t}$. Furthermore,  $\hat{e}_t\coloneqq x_t-\hat{x}_t$ denotes the estimation error at time $t$.
We consider the following cost function 
\begin{align}
&	J(\hat{x}_{t-M_t|t}, \hat{w}_{\cdot|t},  \hat{y}_{\cdot|t},t)=2\eta^{M_t}||\hat{x}_{t-M_t|t}-\hat{x}_{t-M_t}||^2_{P_2} \label{eq:cost}	\\ 
&+\sum_{j=t-M_t}^{t-1} \eta^{t-j-1}2 ||\hat{w}_{j|t}||_Q^2
+\sum_{j\in \mathbb{I}_{[t-M_t,t-1]} \cap K_s} \eta^{t-j-1} ||\hat{y}_{j|t}-y_j||_R^2. \nonumber
\end{align}
The first term of the cost function penalizes the difference between the first element of the estimated state sequence  $\hat{x}_{t-M_t|t}$ and the prior estimate $\hat{x}_{t-M_t}$ that was obtained  at time  $t-M_t$. 
The stage cost then penalizes the estimated noise and the difference between the measured and the estimated output. The weighting matrices $P_2,Q,$ and $R$ and the parameter $\eta$ correspond to parameters in Assumption~\ref{ass:eIOSS}.
If the system is sample-based exponentially i-IOSS,
then the cost function can be parameterized arbitrarily using any positive definite matrices $P_2,Q,$ and $R$
since (\ref{eq:ass1}) can be rescaled accordingly, analogous to the non-sample-based case (cf.  \cite[Remark 1]{Sch23}).
Due to the discount factor $\eta$,  the influence of disturbances and output measurements further in the past is reduced.
\begin{rem}	
Notice that the last summation term in the cost function (\ref{eq:cost}) includes only the output measurements available, and recall that $w$ accounts for both process and measurement noise. 
This implies that the components of the optimal sequence $\hat{w}^*_{\cdot|t}$ that correspond to measurement noise (i.e., that affect only the output equation in (\ref{eq:sys})) at times where there are no measurements (i.e., $j \in \mathbb{I}_{[t-M_t,t-1]}\setminus K_s$) are equal to zero to minimize the cost function. Hence, an equivalent cost function can be formulated by having separate terms for the process and measurement noise, where the term that corresponds to measurement noise would only be included at the time instances  $j\in   \mathbb{I}_{[t-M_t,t-1]}\setminus K_s$, similar to the output summation in (\ref{eq:cost}).
\end{rem}
\par The following proposition shows that if $\delta_t>0$ (meaning no new information is available at time $t$), then, instead of solving (\ref{eq:NLP}), the current state estimate can be equivalently obtained through an open-loop prediction, i.e.,
\begin{align*}
\hat{x}_{t}=f(\hat{x}_{t-1},u_{t-1},0).
\end{align*}
\begin{prop}[{\cite[Proposition 1]{Kra25a}}]
\label{prop:NLPOL}
The solution of the optimization problem  (\ref{eq:NLP}) at time $t\geq0$ is given by 
\begin{align*}
	\begin{aligned}
		\hat{x}^*_{t-M_t|t} &= \hat{x}^*_{t-\delta_t-M_{t-\delta_t}|t-\delta_t},\\
		\hat{w}^*_{j|t} &= \hat{w}^*_{j|t-\delta_t}, \ j\in [t-M_t,t-\delta_t-1],\\
		\hat{w}^*_{j|t} &= 0, \ j \in [t-\delta_t,t-1].
	\end{aligned}
\end{align*}
\end{prop}
The proof can be found in \cite{Kra25a}.
\begin{rem}
\label{rem:2schemes}
The state estimation framework described above requires to solve the MHE problem (\ref{eq:NLP}) when \mbox{$\delta_t=0$}, and to obtain the estimate from a nominal open-loop prediction otherwise.
By Proposition~\ref{prop:NLPOL}, this  is equivalent to solving (\ref{eq:NLP}) at every time instant with a time-varying horizon length $M_t$. Alternatively, we could solve the optimization problem at every time step with a fixed horizon length of $M$ for $t\geq M$ and not only when new information is available. For both approaches RGES  can be established.
The resulting state estimates may differ slightly for the two approaches, but it cannot be said that one method is systematically leading to smaller estimation errors than the other. 
Therefore, we focus here on the approach in which $\hat{x}_t$ is obtained via an open-loop prediction when $\delta_t > 0$, thereby reducing the frequency with which the optimization problem must be solved. 
\end{rem}
\subsection{STABILITY ANALYSIS}
\label{sec:stabana}
In this section, we focus on showing robust stability of the estimation error for the method proposed in Section~\ref{sec:MHEscheme}. 
The  proof of the following theorem uses ideas from \cite{Sch23}, with suitable changes to the here considered sample-based setting.
\begin{thm}[Sample-based MHE is RGES] 
\label{thm:stab}
Consider a set $K$ as in Definition~\ref{def:K}.  
Let Assumption \ref{ass:eIOSS} hold, assume the measurement sequence $K_s \in K$,  and let the horizon $M\in \mathbb{I}_{\geq d_{\text{max}}}$ be chosen such that $4\lambda^2_{\text{max}}(P_2,P_1)\eta^{M}<1$. Then, there exist $\rho \in
[0, 1)$ such that  the state estimation error of the sample-based  MHE
scheme (\ref{eq:NLP}) satisfies for all $t\geq 0$
\begin{align*}
\begin{aligned}
	||\hat{e}_t||&\leq	2 \sqrt{\frac{\lambda_{\text{max}}(P_2,P_1)\lambda_{\text{max}}(P_2)}{\lambda_{\text{min}}(P_1)}}\sqrt{\rho}^{t}||\hat{e}_{0}|| \\&+2\sqrt{\frac{\lambda_{\text{max}}(P_2,P_1)\lambda_{\text{max}}(Q)}{\lambda_{\text{min}}(P_1)}}\sum_{j=0}^{t-1} \sqrt{\rho}^{t-j-1}  ||w_j||.
\end{aligned}
\end{align*}
\end{thm}
\begin{proof}
Due to the NLP constraints (\ref{eq:NLPC1})-(\ref{eq:NLPC4}), the estimated trajectories satisfy (\ref{eq:sys}), $\hat{x}_{j|t}\in \mathbb{X}$ for all $j\in \mathbb{I}_{[t-M_t,t]}$ and $\hat{w}_{j|t}\in \mathbb{W},\hat{y}_{j|t}\in \mathbb{Y}$ for all $j\in \mathbb{I}_{[t-M_t,t-1]}$. Furthermore, we define $\zeta_{\tau}\coloneqq\min\{j|j\in [\tau,\infty)\cap K_s\}-\tau$, i.e., $\zeta_\tau$ is the amount of time steps that will pass until the next measurement from $\tau$ on (and $\zeta_{\tau}=0$ in case that $\tau \in K_s$).
\par Consider some time $t\geq M$. Since $M\geq d_{\text{max}}$, we can  partition the time interval $[t-M_t,t]$ into $[t-M_t,t-M_t+\zeta_{t-M_t}]$ and $[t-M_t+\zeta_{t-M_t},t]$. 
Notice that the second interval is useful because, by definition of $\zeta_{\tau}$, there is a measurement at time $t - M_t + \zeta_{t-M_t}$; hence, every subsequent measurement instance within the second time interval follows the pattern described in Definition~\ref{def:K} using a contiguous subsequence of $\mathcal{D}$.
Thus, we can apply (\ref{eq:ass1}) taking $t-M_t+\zeta_{t-M_t}$ as initial time,  which yields
\begin{align}
\begin{aligned}
	||\hat{x}_t-x_t||_{P_1}^2 &\leq \eta^{M_t-\zeta_{t-M_t}} ||\hat{x}_{t-M_t+\zeta_{t-M_t}|t}^*-x_{t-M_t+\zeta_{t-M_t}}||_{P_2}^2\\&+\sum_{j=t-M_t+\zeta_{t-M_t}}^{t-1}\eta^{t-j-1}||\hat{w}_{j|t}^*-w_j||_Q^2\\& 
	+\sum_{j\in \mathbb{I}_{[t-M_t+\zeta_{t-M_t},t-1]} \cap K_s}\eta^{t-j-1}||\hat{y}_{j|t}^*-y_j||_R^2. 
\end{aligned}
\label{eq:eiossEst}
\end{align}
Due to $\zeta_{t-M_t}\leq d_{\text{max}}$ 
and Assumption~{\ref{ass:eIOSS}} considering all $K_i\in K$ we can again apply (\ref{eq:ass1}) for the time interval $[t-M_t,t-M_t+\zeta_{t-M_t}]$ to obtain
\begin{align}
\begin{aligned}
	{\scalebox{0.94}[1]{$||\hat{x}_{t-M_t+\zeta_{t-M_t}|t}^*-x_{t-M_t+\zeta_{t-M_t}}||_{P_1}^2 \leq\eta^{\zeta_{t-M_t}} ||\hat{x}_{t-M_t|t}^*-x_{t-M_t}||_{P_2}^2$}}\\+\sum_{j=t-M_t}^{t-M_t+\zeta_{t-M_t}-1}\eta^{t-M_t+\zeta_{t-M_t}-j-1}||\hat{w}_{j|t}^*-w_j||_Q^2.
\end{aligned}
\label{eq:eiossEst_eps}
\end{align}
Note that (\ref{eq:eiossEst_eps}) does not contain an output-dependent term since by definition of $\zeta_{t-M_t}$, no measurements were taken in the interval $[t-M_t,t-M_t+\zeta_{t-M_t}-1]$.
Using the fact that
\begin{align*}
\begin{aligned}
	&||\hat{x}_{t-M_t+\zeta_{t-M_t}}-x_{t-M_t+\zeta_{t-M_t}}||_{P_2}^2\\\leq&\lambda_{\text{max}}(P_2,P_1)||\hat{x}_{t-M_t+\zeta_{t-M_t}}-x_{t-M_t+\zeta_{t-M_t}}||_{P_1}^2
\end{aligned}
\end{align*} 
and combining (\ref{eq:eiossEst}) and (\ref{eq:eiossEst_eps}) we can write
\begin{align}
\begin{aligned}
	||\hat{x}_t-x_t||_{P_1}^2 &\leq \lambda_{\text{max}}(P_2,P_1) \big(\eta^{M_t} ||\hat{x}_{t-M_t|t}^*-x_{t-M_t}||_{P_2}^2\\&+\sum_{j=t-M_t}^{t-1}\eta^{t-j-1}||\hat{w}_{j|t}^*-w_j||_Q^2\\& 
	+\sum_{j\in \mathbb{I}_{[t-M_t,t-1]} \cap K_s}\eta^{t-j-1}||\hat{y}_{j|t}^*-y_j||_R^2\big). 
\end{aligned}
\label{eq:eiossEstcomb}
\end{align}
Note that for $t<M$, (\ref{eq:ass1}) is directly applicable, and thus (\ref{eq:eiossEstcomb}) is a valid bound for all $t\geq 0$. 
Since 
\begin{align*}
||\hat{w}_{j|t}^*-w_j||_Q^2\leq 2||\hat{w}_{j|t}^*||_Q^2+ 2||w_j||_Q^2
\end{align*}
and
\begin{align*}
\begin{aligned}
	&||\hat{x}_{t-M_t|t}^*-x_{t-M_t}||_{P_2}^2= ||\hat{x}_{t-M_t}-x_{t-M_t}+\hat{x}_{t-M_t|t}^*-\hat{x}_{t-M_t}||_{P_2}^2\\
	\leq &2||\hat{x}_{t-M_t}-x_{t-M_t}||_{P_2}^2+2||\hat{x}_{t-M_t|t}^*-\hat{x}_{t-M_t}||_{P_2}^2,
\end{aligned}
\end{align*}
we obtain%
\begin{align*}
\begin{aligned}
	||\hat{e}_t||_{P_1}^2&=||\hat{x}_t-x_t||_{P_1}^2 \leq \lambda_{\text{max}}(P_2,P_1)\Big(2 \eta^{M_t} ||\hat{x}_{t-M_t}-x_{t-M_t}||_{P_2}^2\\&+\sum_{j=t-M_t}^{t-1}2\eta^{t-j-1} ||w_j||_Q^2 +J(\hat{x}_{t-M_t|t}^*, \hat{w}_{\cdot|t}^*, \hat{y}_{\cdot|t}^*,t)\Big).
\end{aligned}
\end{align*}
By optimality, we have  $J(\hat{x}_{t-M_t|t}^*, \hat{w}_{\cdot|t}^*, \hat{y}_{\cdot|t}^*,t) \leq J(x_{t-M_t}, w_{\cdot|t}, y_{\cdot|t},t)$ 	with  $w_{\cdot|t}$ and  $y_{\cdot|t}$ referring to the true disturbance and output trajectories on the interval $[t-M_t, t-1]$. Thus,
\begin{align}
\begin{aligned}
	||\hat{e}_t||_{P_1}^2 &\leq \lambda_{\text{max}}(P_2,P_1)\Big(4\eta^{M_t} ||\hat{x}_{t-M_t}-x_{t-M_t}||_{P_2}^2\\&+4\sum_{j=t-M_t}^{t-1}\eta^{t-j-1} ||w_j||_Q^2\Big).
\end{aligned}
\label{eq:boundMtstep1}
\end{align} 
Since $||\hat{x}_{t-M_t}-x_{t-M_t}||_{P_2}^2\leq\lambda_{\text{max}}(P_2,P_1)||\hat{x}_{t-M_t}-x_{t-M_t}||_{P_1}^2$, 
\begin{align*}
\begin{aligned}
	||\hat{e}_t||_{P_1}^2  &\leq 4 \eta^{M_t} \lambda^2_{\text{max}}(P_2,P_1)||\hat{x}_{t-M_t}-x_{t-M_t}||_{P_1}^2\\&+4\lambda_{\text{max}}(P_2,P_1) \sum_{j=t-M_t}^{t-1}\eta^{t-j-1} ||w_j||_Q^2.
\end{aligned}
\end{align*}
Selecting the horizon length  $M$ large enough such that 
\begin{align*}
\rho^M:=4\lambda^2_{\text{max}}(P_2,P_1)\eta^{M}<1
\end{align*}
with $\rho \in [0,1)$, we obtain for all $t\geq M$
\begin{align}
\begin{aligned}
	||\hat{e}_t||_{P_1}^2&\leq \rho^{M_t} ||\hat{e}_{t-M_t}||_{P_1}^2\\&+4\lambda_{\text{max}}(P_2,P_1)\sum_{j=t-M_t}^{t-1}\eta^{t-j-1} ||w_j||_Q^2.
\end{aligned}
\label{eq:boundMstep}
\end{align}
Consider an arbitrary $t\in\mathbb{I}_{\geq M}$ and note that it can be decomposed as $t=l+T$ with $l\in \mathbb{I}_{[0,M-1]}$ and $T\in \mathbb{I}_{\geq M}$ as specified below. 
Using (\ref{eq:boundMtstep1}) we obtain
\begin{align}
\hspace{-0.6em}	
	||\hat{e}_l||_{P_1}^2 &\leq  4 \lambda_{\text{max}}(P_2,P_1)\big(\eta^{l} ||\hat{e}_{0}||_{P_2}^2+\sum_{j=0}^{l-1}\eta^{l-j-1} ||w_j||_Q^2\big).
\label{eq:boundlk}
\end{align}
The  $T$-long time interval consists of $\kappa$ time intervals $[k_{i+1},k_i]$, $i=1,\ldots, \kappa$ with \mbox{$k_{i+1}=k_i-M-\delta_{k_i}$} and $k_{1}=t$. Recall $\delta_{k_i}=k_i-1-\max\{0,j \in K_s|j<k_i\}$.
Applying (\ref{eq:boundMstep})  for each of the $\kappa$ time intervals and using $\eta \leq \rho $ yields
\begin{align*}
\begin{aligned}
	||\hat{e}_{k_i}||_{P_1}^2&\leq \rho^{M+\delta_{k_i}} ||\hat{e}_{k_{i+1}}||_{P_1}^2\\&+4\lambda_{\text{max}}(P_2,P_1)\sum_{j=k_i-M-\delta_{k_i}}^{k_i-1}\rho^{k_i-j-1} ||w_j||_Q^2.
\end{aligned}
\end{align*}
Applying this inequality recursively we derive the following upper bound for the estimation error
\begin{align*}
\begin{aligned}
	&||\hat{e}_t||_{P_1}^2 \leq \rho^{T}	||\hat{e}_l||_{P_1}^2+4\lambda_{\text{max}}(P_2,P_1)\sum_{j=l}^{t-1}\rho^{t-j-1}||w_j||_Q^2 \\
	&\stackrel{(\ref{eq:boundlk})}{\leq} 4\lambda_{\text{max}}(P_2,P_1)\rho^{T}\Big(\eta^{l} ||\hat{e}_{0}||_{P_2}^2+\sum_{j=0}^{l-1}\eta^{l-j-1} ||w_j||_Q^2\Big) \\&+4\lambda_{\text{max}}(P_2,P_1)\sum_{j=l}^{t-1}\rho^{t-j-1}||w_j||_Q^2.
\end{aligned}
\end{align*}
Due to $\eta\leq \rho$ and $t=l+T$ we can write
\begin{align*}
\begin{aligned}
	||\hat{e}_t||_{P_1}^2&\leq  4 \lambda_{\text{max}}(P_2,P_1)\big(\rho^{t}||\hat{e}_{0}||_{P_2}^2+\sum_{j=0}^{t-1} \rho^{t-j-1}  ||w_j||_Q^2\big).
\end{aligned}
\end{align*}
Note that $\lambda_{\text{min}}(P_1)||\hat{e}_t||^2\leq ||\hat{e}_t||_{P_1}^2$, $||\hat{e}_0||_{P_2}^2\leq \lambda_{\text{max}}(P_2)||\hat{e}_0||^2$ and $||w_j||_{Q}^2\leq \lambda_{\text{max}}(Q)||w_j||^2$. Thus, using this and the fact that  $\sqrt{a+b}\leq \sqrt{a}+\sqrt{b}$ for all $a,b\geq0$ we obtain
\begin{align}
\begin{aligned}
	||\hat{e}_t||&\leq	2 \sqrt{\frac{\lambda_{\text{max}}(P_2,P_1)\lambda_{\text{max}}(P_2)}{\lambda_{\text{min}}(P_1)}}\sqrt{\rho}^{t}||\hat{e}_{0}|| \\&+2\sqrt{\frac{\lambda_{\text{max}}(P_2,P_1)\lambda_{\text{max}}(Q)}{\lambda_{\text{min}}(P_1)}}\sum_{j=0}^{t-1} \sqrt{\rho}^{t-j-1}  ||w_j||.
\end{aligned}
\label{eq:boundsum}
\end{align}
For the case $t<M$, (\ref{eq:boundsum}) can be obtained directly from (\ref{eq:boundMtstep1}). We conclude that (\ref{eq:boundsum}) holds for all $t \geq 0$, completing the proof.
\end{proof}
\begin{rem} 
The stability proof can be performed in a similar manner for the modified sample-based MHE scheme discussed  in Remark~\ref{rem:2schemes}.
\end{rem}
\section{LINEAR SYSTEMS AND SAMPLE-BASED  i-IOSS}
\label{sec:LinSys}
In the following we will analyze the connections between sample-based observability and sample-based i-IOSS for linear systems. We show that, in the linear case, sample-based observability of the unstable subsystem implies sample-based exponential i-IOSS of the system. Hence, previously established conditions for linear systems to be sample-based observable (cf. \cite{Kra22}) can be utilized to verify or design sampling strategies that satisfy Assumption~\ref{ass:eIOSS} to ensure RGES of the sample-based MHE.
\par Consider a linear time-invariant system described by
\begin{align}
\begin{aligned}
x_{t+1}&=Ax_t+B u_t+w_t,\\
y_t&=C x_t +D u_t
\end{aligned}
\label{eq:syslin}
\end{align}
where $A \in \mathbb{R}^{n \times n}$, $B \in \mathbb{R}^{n \times m}$, $C \in \mathbb{R}^{p \times n}$, $D \in \mathbb{R}^{p \times m}$. 
For convenience in deriving the results later in this section, we do not consider measurement noise. 
This is without loss of generality, as sample-based i-IOSS of system (\ref{eq:syslin}) is equivalent to sample-based i-IOSS of system (\ref{eq:syslin}) with the output equation  perturbed by measurement noise, i.e., 	$y_t=C x_t +D u_t +v_t$, $v_t \in \mathbb{R}^p$.
\par In the non-sampled case, it is well known that for linear systems, exponential i-IOSS is equivalent to detectability \cite{Son95,Knu20}. 
In the following, we analyze the relationship between sample-based observability and sample-based i-IOSS. For this we need the following definitions of the sample-based observability matrix and sample-based observability.
\begin{defi}[Sample-based observability matrix \cite{Kra25b}]
\label{def:SOm}
Consider a set of arbitrary time instances $\{ \tau_i \}_{i=1}^k$ for some $k\geq 1$, with $0 \leq \tau_1$ and $\tau_i < \tau_{i+1}$ for all $i=1,...,k-1$.
The sample-based observability matrix  is given by
\begin{align*}
O_{\text{s}}(A,C)=\begin{pmatrix}
	(CA^{\tau_1})^\top&(CA^{\tau_2})^\top& \ldots&(CA^{\tau_k})^\top 
\end{pmatrix}^\top.
\end{align*}
\end{defi}
\begin{defi}[Sample-based observability \cite{Kra25b}]
\label{def:SO}
System~(\ref{eq:syslin}) is  sample-based  observable for a given sampling sequence 	$\{ \tau_i \}_{i=1}^k$ for some $k\geq 1$, with $0 \leq \tau_1$ and $\tau_i < \tau_{i+1}$ for all $i=1,\ldots,k-1$,  if for any initial state $x_0$ and input $u_t$, \color{new} and no disturbances (i.e.,  $w(t)=0, \forall t\geq 0$), \color{black} the value of $x_0$ can be uniquely reconstructed from the knowledge of the input trajectory on the interval $[0,\tau_k]$ and the sampled  outputs $y_{\tau_i}$, $i=1,\ldots,k$, 
i.e., the sample-based observability matrix $O_{\text{s}}$ in Definition \ref{def:SOm} has full column rank.
\end{defi}
\par In the following we want to show that if there exists a finite time $T$  such that sample-based observability holds over every interval of length $T$, then this implies sample-based exponential \mbox{i-IOSS} according to Assumption~\ref{ass:eIOSS}.
To prove the corresponding theorem, we rely on a sufficient condition for sample-based exponential \mbox{i-IOSS}. The condition was originally stated in \cite{Kra25} for general nonlinear asymptotically i-IOSS systems, and its theorem and proof can be straightforwardly adapted to exponentially i-IOSS systems. In the following, we state this theorem directly for (\ref{eq:syslin}).
\begin{thm}[{\scalebox{0.95}[1]{Sufficient condition sample-based i-IOSS \cite{Kra25}}}]
Consider system (\ref{eq:syslin}) and let $(A,C)$ be detectable.
Moreover, consider  some set $K$ according to Definition~\ref{def:K}. Then, the system is sample-based exponentially i-IOSS with respect to $K$ if there exist some constants $a_w, a_h > 0$ and a finite time $t^*$ such that for any two initial conditions $x_{0}$, $\tilde{x}_{0}$, any two input (disturbance) trajectories $\mathbf{w},\mathbf{\tilde{w}} \in\mathbb{W}^\infty$ and all $u\in\mathbb{U}^\infty$ the following holds  for all $t\geq t^*$ and for all $K_i \in K$
\begin{align}
	||\Delta y_t||\leq &\max\{ \max_{ j\in \mathbb{I}_{[0,t-1]} \cap K_i} a_h|| \Delta y_{j}||,\max_{j\in \mathbb{I}_{[0,t-1]}} a_w|| \Delta w_{j}		||\}
\label{eq:cond_sampling}
\end{align}
with $\Delta w_t=w_{t}-\tilde{w}_t$ and $\Delta y_t=C \Delta x_t=Cx_{t}-C\tilde{x}_t$.
\label{thm1}
\end{thm}
\par Using this sufficient condition, we can now establish the following theorem linking sample-based observability to  sample-based i-IOSS.
\begin{thm}
Consider a set $K$ as in Definition~\ref{def:K}.
If there exists a finite $T$ such that the sample-based observability matrix with time indices in the set   $K_1 \cap [t-T-1,t-1]$ has full column rank for all $t>T$, then the system (\ref{eq:syslin}) is sample-based exponentially i-IOSS with respect to $K$.
\label{thm:sbobsviIOSS}
\end{thm}
\begin{proof}
Since the sample-based observability matrix has full column rank, the system is sample-based observable, and therefore also observable. This in turn implies that the system is exponentially i-IOSS (cf. proof of Theorem 6 in \cite{Knu20}). 
Hence, it only remains to be shown that (\ref{eq:cond_sampling}) holds. 
Consider the output  difference at some time $t> T$
\begin{align}
\label{eq:Deltay}
\Delta y_t=C\Delta x_t=CA^{T+1}\Delta x_{t-T-1}+\sum_{j=t-T-1}^{t-1} CA^{t-1-j}\Delta w_j.
\end{align}
Next we want to replace $CA^{T+1}$. For this,  consider the sampling set $K_1$ and let $\tau_{1,t}+t-T-1, \tau_{2,t}+t-T-1,\dots,\tau_{k_t,t}+t-T-1$  be the $k_t$ time  instances in  $K_1 \cap [t-T-1,t-1]$. 
Now, the sample-based observability matrix corresponding to this interval can be written as
\begin{align}
O_{\text{s},t}(A,C)=\begin{pmatrix} CA^{\tau_{1,t}}\\CA^{\tau_{2,t}}\\ \vdots \\ CA^{\tau_{k_t,t}}\end{pmatrix}A^{t-T-1}
\end{align}
and, since $O_{\text{s},t}$ has full column rank by assumption, it also holds that
\begin{align}
\mathtt{rank}\begin{pmatrix} (CA^{\tau_{1,t}})^\top& \ldots & (CA^{\tau_{k_t,t}})^\top\end{pmatrix}=n.
\label{eq:sbObsv}
\end{align}
Hence, there exist matrices $\alpha_{j,t} \in \mathbb{R}^{p\times p}$  such that we can write
\begin{align}
CA^{T+1} = \sum_{j=1}^{k_t} \alpha_{j,t} CA^{\tau_{j,t}}.   
\label{eq:CAT+1}
\end{align}
We now show that there exists a constant $\bar\alpha >0$ such that $\| \alpha_{j,t} \| \leq \bar\alpha$ for all $j \in [1,k_t]$ and all $t>T$. Notice that the exponents in (\ref{eq:sbObsv}) satisfy $\{\tau_{j,t}\}_{j=1}^{k_t} \subseteq \{0,1,\dots,T\}$ for any $t>T$, i.e., the matrix in (\ref{eq:sbObsv}) is always a subset of the rows of the extended observability matrix $\bar{O}(A,C)= \begin{pmatrix}C^\top& (CA)^\top & \ldots & (CA^T)^\top\end{pmatrix}^\top$. Clearly, there are only a finite set of possible submatrices that can be formed using the block rows of $\bar{O}(A,C)$. In fact, there are at most
\begin{align*}
\sum_{r=\nu}^{T+1}\Omega_r, \quad \Omega_r=\binom{T+1}{r}
\end{align*}	
different possible submatrices of $\bar{O}$ that satisfy (\ref{eq:sbObsv}), where $\nu$ is the observability index of system  (\ref{eq:syslin}). For each of these submatrices we can define a set of matrices $\alpha_{j,t}$, and finally select an $\bar\alpha$ such that $\| \alpha_{j,t} \| \leq \bar\alpha$ for all $j \in [1,k_t]$ and all $t>T$ as desired.
Thus, from (\ref{eq:Deltay}) and (\ref{eq:CAT+1}), we get
\begin{align*}
\begin{aligned}
	&	||\Delta y_t||\leq\sum_{j\in [t-T-1,t-1] \cap K_1 } \bar{\alpha} ||CA^{j-t+T+1}\Delta x_{t-T-1}||\\&+\sum_{j=t-T-1}^{t-1} ||CA^{t-1-j}\Delta w_j||
	\leq \sum_{j\in [t-T-1,t-1] \cap K_1 } \bar{\alpha} \Big(||\Delta y_{j}||\\&+\sum_{i=t-T-1}^{j-1}||CA^{j-1-i}\Delta w_i||\Big)+\sum_{j=t-T-1}^{t-1}|| CA^{t-1-j}|| \ ||\Delta w_j||
\end{aligned}
\end{align*}
where we used that $\Delta y_{j}=CA^{j-t+T+1}\Delta x_{t-T-1}+\sum_{i=t-T-1}^{j-1}CA^{j-1-i}\Delta w_i$.
We can bound $||CA^j||\leq k_w$ for all $j\in[0,T]$ with some $k_w>0$. Hence, we obtain the following upper bound for all $t>T$
\begin{align}
\begin{aligned}
	||\Delta y_t||&\leq\sum_{j\in [0,t-1]\cap K_1} \bar{\alpha}  ||\Delta y_{j}||+\sum_{j=0}^{t-1} c_w ||\Delta w_j||	
\end{aligned}
\label{eq:sufconK1}
\end{align}
with $c_w=k_w (\bar{\alpha}(T+1)+1)$. Since we assume sample-based observability in any $T$-long interval, (\ref{eq:sufconK1}) also holds for all $K_i\in K$ by definition of $K_i$ and $K$ in Definition~\ref{def:K}.
Thus selecting $a_h=\bar{\alpha} $, $a_w=c_w $ and $t^*=T+1$ satisfies (\ref{eq:cond_sampling}) and completes the proof.
\end{proof}
\par 	We can relax the result from Theorem~\ref{thm:sbobsviIOSS}. That is, in the following we show that sample-based observability of only the part of the system that is not asymptotically stable is sufficient for  sample-based exponential  i-IOSS. This is formulated in the theorem below.
For this, consider the system in Jordan canonical form with
\begin{align*}
A_{\text{J}}=T_{\text{J}}^{-1}AT_{\text{J}}=\begin{pmatrix}
A_{\text{s}}&0\\0&A_{\text{us}} \end{pmatrix}, \quad C_{\text{J}}=CT_{\text{J}}=\begin{pmatrix} C_{\text{s}}&C_{\text{us}}\end{pmatrix}
\end{align*}
where $A_{\text{s}}$ contains all eigenvalues of $A$ that lie strictly inside the unit disc and $A_{\text{us}}$ all the remaining ones (i.e., those located on or outside the unit disc),
$x^{\text{J}}=T_{\text{J}}^{-1}x=\begin{pmatrix}x^{\text{s}^\top} &x^{\text{us}^\top}\end{pmatrix}^\top$ and $w^{\text{J}}=T_{\text{J}}^{-1}w=\begin{pmatrix}w^{\text{s}^\top} &w^{\text{us}^\top}\end{pmatrix}^\top$ with the vectors $x^{\text{s}}$, $x^{\text{us}}$ and $w^{\text{s}}$, $w^{\text{us}}$ following the partition of $A_{\text{J}}$ above. Additionally, we denote $y_t^{\text{s}}\coloneq C_{\text{s}}x_t^{\text{s}}$ and $y_t^{\text{us}}\coloneq C_{\text{us}}x_t^{\text{us}}$.
\begin{thm}
\label{thm:usSBobsv}
Consider a set $K$ as in Definition~\ref{def:K}.
If there exists a finite $T$ such that the sample-based observability matrix  of the pair $(A_{\text{us}},C_{\text{us}})$ with time indices in the set   $K_1 \cap [t-T-1,t-1]$ has full column rank for all $t>T$, then the system (\ref{eq:syslin}) is sample-based exponentially i-IOSS with respect to $K$.
\end{thm}
\begin{proof}
Recall from Remark~\ref{rem:eIOSS} that the sum-based formulation of sample-based exponential i-IOSS is equivalent to a max-based formulation analogous to (\ref{eq:td-iIOSS}). Hence, applying Theorem~\ref{thm:sbobsviIOSS} to the subsystem $(A_{\text{us}},C_{\text{us}})$ yields for all $t\geq 0$ and for all $K_i\in K$
\begin{align}
||\Delta x^{\text{us}}_t||&\leq \max\Big\{a_\text{us}||\Delta x^{\text{us}}_0||\lambda^t,\max_{j\in\mathbb{I}_{[0,t-1]}}b_\text{us}||\Delta w^{\text{us}}_j||\lambda^{t-j-1}, \nonumber \\
& \max_{j\in\mathbb{I}_{[0,t-1]}\cap K_i}c_\text{us}||\Delta y^{\text{us}}_j||\lambda^{t-j-1}\Big\} 	
\label{eq:iIOSSinstab}
\end{align}
with  $a_{\text{us}},b_{\text{us}},c_{\text{us}} \in \mathbb{R}_{>0}$ and $\lambda \in(0,1)$.
Since $A_{\text{s}}$ is a stable matrix and hence the corresponding subsystem is i-ISS, it holds for all $t\geq 0$ that
\begin{align}
	||\Delta x^{\text{s}}_t||&\leq \max\Big\{a_{\text{s}}||\Delta x^{\text{s}}_0||\lambda^t,\max_{j\in\mathbb{I}_{[0,t-1]}}b_{\text{s}}||\Delta w^{\text{s}}_j||\lambda^{t-j-1}\Big\}
\label{eq:stabiISS}
\end{align}
with  $a_{\text{s}},b_{\text{s}} \in \mathbb{R}_{>0}$.
Note that we consider here the same $\lambda$ in both (\ref{eq:iIOSSinstab}), and (\ref{eq:stabiISS}) which is always possible by selecting the maximum of the values required by either bound. 
Due to $||\Delta x^{\text{J}}_t||\leq  \max\{\sqrt{2}||\Delta x^{\text{us}}_t||,\sqrt{2}||\Delta x^{\text{s}}_t||\}$
and $||\Delta y_t^{\text{us}}||=||\Delta y_t^{\text{us}}+\Delta y_t^{\text{s}}-\Delta y_t^{\text{s}}||\leq ||\Delta y_t||+||\Delta y_t^{\text{s}}||$ we can write for all $t\geq 0$ and  all $K_i\in K$ 
\begin{align*}
\begin{aligned}
	||\Delta x^{\text{J}}_t||&\leq \sqrt{2} \max\Big\{a_{\text{us}}||\Delta x^{\text{us}}_0||\lambda^t,\max_{j\in\mathbb{I}_{[0,t-1]}}b_{\text{us}}||\Delta w^{\text{us}}_j||\lambda^{t-j-1},\\
	& \max_{j\in\mathbb{I}_{[0,t-1]}\cap K_i}c_{\text{us}}(||\Delta y_j||+||\Delta y^{\text{s}}_j||)\lambda^{t-j-1},\\& 
	a_{\text{s}}||\Delta x^{\text{s}}_0||\lambda^t,\max_{j\in\mathbb{I}_{[0,t-1]}}b_{\text{s}}||\Delta w^{\text{s}}_j||\lambda^{t-j-1}\Big\}.
\end{aligned}
\end{align*}
Since \begin{align*}
\begin{aligned}||\Delta y^{\text{s}}_j||\leq ||C_{\text{s}}|| \ ||\Delta x^{\text{s}}_j|| &\leq \max\Big\{a_{\text{s}}||C_{\text{s}}|| \ ||\Delta x^{\text{s}}_0||\lambda^j,\\&\max_{i\in \mathbb{I}_{[0,j-1]}} b_{\text{s}}||C_{\text{s}}||\ ||\Delta w^{\text{s}}_i||\lambda^{j-i-1}\Big\}
\end{aligned}
\end{align*}
we obtain for all $t\geq 0$ and  all $K_i\in K$ 
\begin{align}
	||\Delta x^{\text{J}}_t||&\leq \sqrt{2} \max\Big\{a_{\text{us}}||\Delta x^{\text{us}}_0||\lambda^t,\max_{j\in\mathbb{I}_{[0,t-1]}}b_{\text{us}}||\Delta w^{\text{us}}_j||\lambda^{t-j-1},\nonumber\\
	& \max_{j\in\mathbb{I}_{[0,t-1]}\cap K_i}2c_{\text{us}}||\Delta y_j|| \lambda^{t-j-1},2c_{\text{us}} a_{\text{s}}||C_{\text{s}}||\ ||\Delta x^{\text{s}}_0||\lambda^{t-1},\nonumber\\&\max_{j\in\mathbb{I}_{[0,t-1]}}
	2 c_{\text{us}} b_{\text{s}}||C_{\text{s}}||\ ||\Delta w^{\text{s}}_j||\lambda^{t-j-1}, \nonumber \\
	&a_{\text{s}}||\Delta x^{\text{s}}_0||\lambda^t,\max_{j\in\mathbb{I}_{[0,t-1]}}b_{\text{s}}||\Delta w^{\text{s}}_j||\lambda^{t-j-1}\Big\} 	\label{eq:sbiIOSSJordan}
	\\
	&\leq \sqrt{2}\max\Big\{\max\{a_{\text{us}},a_{\text{s}},2c_{\text{us}} a_{\text{s}}||C_{\text{s}}||\lambda^{-1}\}||\Delta x^{\text{J}}_0||\lambda^t,\nonumber\\
	&\max_{j\in\mathbb{I}_{[0,t-1]}}\max\{b_{\text{us}},2 c_{\text{us}} b_{\text{s}}||C_{\text{s}}||,b_{\text{s}}\} ||\Delta w_j^{\text{J}}||\lambda^{t-j-1},\nonumber\\
	&\max_{j\in\mathbb{I}_{[0,t-1]}\cap K_i}2c_{\text{us}}||\Delta y_j|| \lambda^{t-j-1}\Big\}.\nonumber
\end{align}
Applying $\sigma_{\text{min}}(T_{\text{J}}^{-1})||\Delta x_t||\leq||\Delta x^{\text{J}}_t||$, $||\Delta x^{\text{J}}_0||\leq \sigma_{\text{max}}(T_{\text{J}}^{-1})||\Delta x_0||$ and $||\Delta w^{\text{J}}_t||\leq \sigma_{\text{max}}(T_{\text{J}}^{-1})||\Delta w_t||$ to (\ref{eq:sbiIOSSJordan})  
yields for all $t\geq 0$ and all $K_i\in K$ 
\begin{align*}
	\begin{aligned}
		&||\Delta x_t||
		\leq \frac{\sqrt{2}}{\sigma_{\text{min}}(T_{\text{J}}^{-1})}\max\Big\{
		\\&\max\{a_{\text{us}},a_{\text{s}},2c_{\text{us}} a_{\text{s}}||C_{\text{s}}||\lambda^{-1}\}\sigma_{\text{max}}(T_{\text{J}}^{-1})||\Delta x_0||\lambda^t,\\
		&\max_{j\in\mathbb{I}_{[0,t-1]}}\max\{b_{\text{us}},2 c_{\text{us}} b_{\text{s}}||C_{\text{s}}||,b_{\text{s}}\}\sigma_{\text{max}}(T_{\text{J}}^{-1}) ||\Delta w_j||\lambda^{t-j-1},\\
		&\max_{j\in\mathbb{I}_{[0,t-1]}\cap K_i}2c_{\text{us}}||\Delta y_j|| \lambda^{t-j-1}\Big\},
	\end{aligned}
\end{align*}
i.e., the sample-based i-IOSS bound. As discussed in Remark~\ref{rem:eIOSS}, this is equivalent to a sum-based bound as in (\ref{eq:ass1}), thus concluding the proof.				
\end{proof}
\section{NUMERICAL EXAMPLE}
\label{sec:numex}
\begin{figure}[!t]
\centering
\centerline{\input{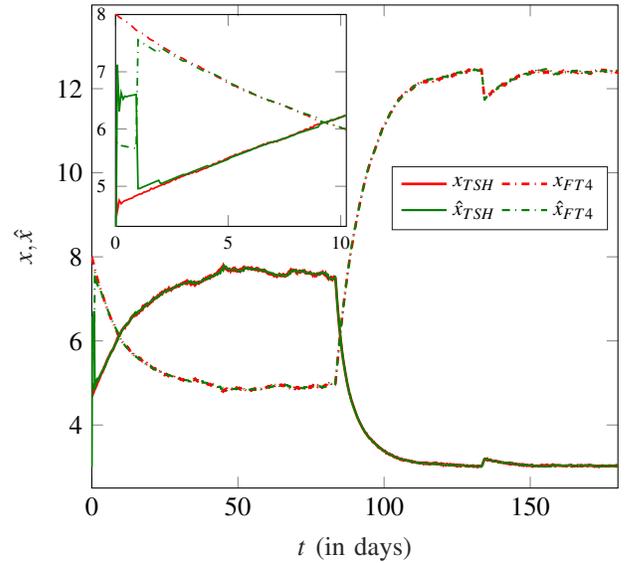}}
\caption{True states (red) and sample-based MHE results (green) when considering a measurement once a day, i.e., on average every twelfth time instant a measurement was taken. In the top-left, a zoomed-in plot of the first 10 days.} 
\label{fig:simmhe}
\end{figure}

\begin{figure}[!t]
\centering
\centerline{\input{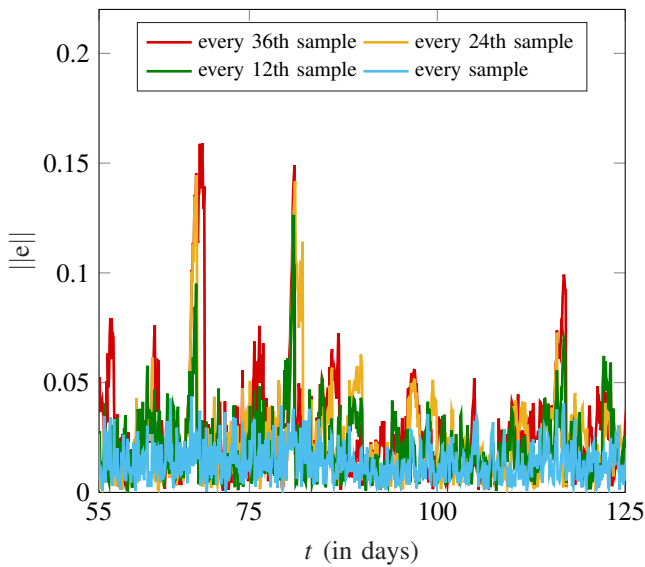}}
\caption{Estimation error  for four different settings. We simulated the cases that a measurement was taken  12 times a day (blue), i.e., at every time instant, 
	once a day (green), every second day (yellow),  and every third day (red).} 
\label{fig:simer}
\vspace{-1.2em}
\end{figure}

Consider the simplified two-dimensional model of the hypothalamic–pituitary–thyroid axis from \cite{Yan21}, which describes hormone dynamics relevant to the study of thyroid disorders. In such biomedical applications, obtaining  measurements typically requires taking blood samples,
which is impractical to do on a frequent basis. Thus, in the following we apply the proposed MHE scheme in Section~\ref{sec:MHEscheme} to reconstruct the internal states of the system using only infrequent and irregular measurements. We discretized the system using the Euler method with a sampling time of  $\tau=2\mathtt{h}$ to obtain the following system model	
\begin{align*}
\begin{aligned}
	x_{TSH,t+1}&=\frac{p_1\tau(U-x_{FT4}(t))}{s_1+x_{FT4}(t)}+p_1\tau+(1-d_1\tau)x_{TSH,t}+w_{1,t}\\
	x_{FT4,t+1}&=\frac{\tau p_2 x_{TSH,t}}{s_2+x_{TSH,t}}		+(1-d_2\tau)x_{FT4}(t)+G\tau+w_{2,t}
\end{aligned}
\end{align*}
with state $x=\begin{bmatrix}x_{TSH}, & x_{FT4}\end{bmatrix}^\top$ where $x_{TSH}$ and $x_{FT4}$ refer to the concentrations of two thyroid hormones, thyroid-stimulating hormone (TSH) and free thyroxine (FT4), respectively, with $U$ denoting the euthyroid 
set point of $FT4$, and $G$ representing the amount of
increased or decreased FT4 levels due to the  intake of thyroid drugs. 
The description of the parameters $p_1,p_2,s_1,s_2$ can be found in \cite{Yan21}. We choose $p_2$ such  that the case of  deficient production of thyroid hormones is modeled, which is called hypothyroidism. 
We assume $x_{TSH}$ as our measured output, i.e, \mbox{$y_t=\begin{bmatrix} 1 &0 \end{bmatrix} x_t+w_{3,t}$}.
We consider that the system is affected by disturbances $w_{1,t},w_{2,t},w_{3,t}$. In
the simulation, the additive disturbance $w \in \mathbb{R}^3$ is treated as
a uniformly distributed random variable that satisfies $|w_i| \leq
0.01$, $i = 1,2,3$.
We consider an irregular measurement sequence where on average every twelfth  sample is available, i.e., a measurement was taken once per day.  We apply the sample-based MHE scheme from Section~\ref{sec:MHEscheme} with a horizon length $M=100$ to estimate the states (cf. Figure~\ref{fig:simmhe}) over a simulation of 180 days. \color{new} The cost function is parameterized using $\eta=0.91$, $R=1.57$,
\begin{align*}
P_2=\begin{bmatrix}1.09&7.32\\7.32&53.94\end{bmatrix}, \ 	Q=\begin{bmatrix}104.94&0&0\\0&773.78&0\\0&0&1567.8\end{bmatrix}.
\end{align*}
\color{black}
We consider \mbox{$x_0=[5,8]^\top$} and the initial estimate $\hat{x}_0=[3,6]^\top$.
We first simulate an untreated case of hypothyroidism, after approximately 83 days we simulate the intake  of daily medication by setting $G=0.75$ such that eventually the euthyroid set point is reached\footnote{The simulations were performed in Matlab with the NLP solver IPOPT.  Matlab code is available at  https://doi.org/10.25835/a69nkad5.}. 
In our simulation, we consider the case where the patient does not take the medication once, at day 130. The result is shown in Figure~\ref{fig:simmhe}.
It can be observed that the trajectories of the estimated state closely follow the true state trajectories in all considered settings (medicated, no intake of medication).
Next, we simulate the same system but using four different cases for the sampling scheme. Namely, we consider the cases of having (i) a measurement at every time instant (i.e., standard MHE for comparison), and irregular measurement schemes with on average one measurement (ii) per day, (iii) every second day, and (iv) every third day. 
The  estimation errors for a representative segment  (day 55 to day 125) of the 180-day simulation are shown in Figure~\ref{fig:simer}. 
The estimation error converges to a neighborhood of zero. As expected, the error is smaller when more measurements are available for the sample-based MHE scheme. This is also reflected in Table~\ref{tab:er}, which shows the root mean squared error (RMSE) over the complete simulation spanning 180 days. The table again includes results for measurements taken at every time instant, on average at every 12th, 24th, and 36th time instant, and additionally for measurements taken on average at every 6th time instant (i.e., twice a day).
\begin{table}[h]
\caption{RMSE for different measurement sequences.}
\label{tab:er}
\centering
\begin{tabular}{l c}
\hline
\rule{0pt}{2.5ex} Measurements on average & RMSE \\
\hline
\rule{0pt}{2.5ex}every two hours (i.e. every measurement) &  0.1105         
\\
twice a day &  0.1803  \\
once a day &  0.2383  \\
every second day &  0.3068 \\
every third day &  0.3622 \\
\hline
\end{tabular}
\end{table}
\section{CONCLUSION}
In this paper, we have developed a  sample-based moving horizon estimation scheme designed for systems with irregularly and/or infrequently available measurements. The cost function of the MHE optimization problem is adapted to account for irregular sampling. When no new measurements are available, the optimization problem does not need to be  solved. 
Moreover, we showed that, under the assumption of sample-based exponential i-IOSS,  the estimator guarantees robust global exponential stability. Additionally, by establishing connections between sample-based observability and i-IOSS for linear systems, we provide a means to verify or design sampling strategies that ensure the detectability  assumption required to guarantee RGES. The effectiveness of the proposed approach was demonstrated through a simulation example of a biomedical application. %

\end{document}